# Drop Spray Electrification

Riko Kazama[1], Masaya Toda[1], Hans-Jürgen Butt[2,*], and Rutvik Lathia[2,*]

[1] Graduate School of Engineering, Tohoku University, Sendai, Miyagi, Japan

[2] Max Planck Institute for Polymer Research, Mainz, Germany

*Corresponding authors: butt@mpip-mainz.mpg.de; lathiar@mpip-mainz.mpg.de

# Abstract

Charge separation during the breakup of water drop has been recognized since the early studies of waterfall and spray electrification, yet the role of controlled drop fragmentation at structured liquid-repellent surfaces remains unclear. Here we show that water drops impacting superhydrophobic meshes generate charged secondary droplets as liquid penetrates and fragments through the mesh pores. By combining Faraday-cup charge measurements with high-speed imaging, we identify how the charge depends on the Weber number and on the pathway of spray formation. No measurable charge is detected below the penetration threshold. At low Weber numbers, recoil jet formation gives a high charge per unit spray mass, whereas at intermediate Weber numbers both recoil jets and impact-induced jets contribute to charging. At higher Weber numbers, pancake bouncing suppresses recoil jet formation, so charging is dominated by impact induced penetration and the total charge approaches a saturated value. We further show that smaller mesh pores enhance the charge-to-mass ratio and that conductive meshes provide stable charging under repeated impacts by dissipating residual surface charge. These findings provide design principles for superhydrophobic mesh platforms for spray charging and rain-driven energy harvesting.

# I. INTRODUCTION

Over the last decade, it has become apparent that electric energy can be generated by drops sliding down a tilted plate or by water flowing through capillaries. The gravitational energy of the liquid moving from the top of the plate or capillary to its lower end is converted into electrical energy. This phenomenon has stimulated the development of different types of triboelectric nanogenerators (TENGs).[1-5] The generation of electric energy from sliding drops is interesting because it does not require any moving parts and could be used to power remote sensors, for example.

However, the gravitational energy of drops moving down an inclined plate is limited. As an alternative, we were motivated to use the kinetic energy of falling water drops, e.g. in rain, and convert it into electricity. In addition to sliding, the splitting of drops can also lead to charge separation. In 1892, Lenard observed that small fragments of water produced in a splash carry negative charges, while large fragments carry positive charges.[6] Thomson noted that water droplets tend to be positively charged when they splash off a solid surface[7]. Simpson and Nolan hit a falling drop of distilled water by a jet of air.[8,9] The air jet split the primary drop into smaller droplets and they found that the small droplets carry charges. Lenard suggested that negative and positive layers of charge are present near the water-air interface, and that charges are separated when a mass of water breaks up as a result of the disruption of this layer.[10] Darvish demonstrated that when a water drop moves across an array of Janus micropillar, tiny satellite droplets form on each micropillar.[11] Each of these satellite droplets is charged, generating the opposite charge in the primary drop. Using arrays of Janus micropillars enables satellite droplets to be split off in a very defined way.

To separate charges from falling drops, we are exploring the potential of using superhydrophobic coated meshes. It is known that, when impacting superhydrophobic meshes, water droplets tend to form a spray of tiny secondary droplets if they have a sufficiently high impact velocity.[12-16]

Building on these earlier observations of charge separation during drop breakup and splashing, we ask whether a porous mesh could be used as a simple passive platform to separate the charge from an impacting water drop, converting it into a spray of charged secondary droplets. In particular, we investigate whether charge generation is solely dependent on the amount of liquid ejected through the mesh, or if distinct spray-formation dynamics play a role. We show that the spray carries electric charge and that the amount of charge depends on the impact velocity, mesh geometry, and droplet dynamics. By combining charge measurements with high-speed imaging, we identify the physical regimes responsible for spray charging and establish design principles for improving charge separation using porous superhydrophobic surfaces. This approach allows us to identify the physical mechanisms responsible for mesh-induced spray charging and to evaluate the potential of superhydrophobic meshes for harvesting charge from falling drops such as rain.

# II. MATERIALS AND METHODS

## a. Superhydrophobic mesh fabrication

Steel mesh (400 μm pore size, $w$, and 330 μm wire diameter, $d_w$) was sourced locally from Peenya, Bangaluru, India. Copper meshes ($w$ = 533 μm, 285 μm and 133 μm pore sizes with $d_w$ = 300 μm, 225 μm and 110 μm wire diameter, respectively) were purchased from TWP inc, USA. To try an electrically insulating material, polypropylene meshes ($w$ = 500 μm and $d_w$ = 300 μm) were purchased from neoLab. Tetraethoxysilane (TEOS, 99+%, thermo scientific), aqueous ammonia (25%, VWR chem.) and trichloro(perfluorooctyl)silane (97%, abcr GmbH), trichloromethylsilane (99%, Sigma-Aldrich), perfluorodecyltrichlorosilane (97%, abcr GmbH) were used for hydrophobization.

Steel and copper mesh were used to make superhydrophobic surface as described previously.[17,18] Meshes were sequentially cleaned in an ultrasonic bath containing deionized water, ethanol, and deionized water for 10 min in each solvent to remove surface contaminants. The cleaned meshes were then dried in a vacuum oven at 60 °C for 2 h and subjected to oxygen plasma at 100% plasma power (Diener electronic, with 300 W and 0.3 bar) for 5 min to activate the surface. A sacrificial carbon soot layer was deposited by holding each mesh in the candle flame for 90 s to achieve uniform surface coverage. The soot-coated meshes were placed in a sealed desiccator containing two separate beakers, one filled with 3 mL aqueous ammonia ($NH_3$) and the other with 3 mL TEOS, and kept overnight at room temperature to enable vapor-phase silica deposition through the hydrolysis and condensation of TEOS. Subsequently, the meshes were calcined in a furnace at 600 °C for 5 h to remove the carbon soot template and form a porous silica coating. To impart low surface energy and achieve superhydrophobicity, the calcined meshes were transferred to a second desiccator containing a beaker with 0.2 mL trichloro(perfluorooctyl)silane and exposed to its vapor under a reduced pressure of 45 mbar for 2 h. Finally, the desiccator was vented to atmospheric pressure, and the functionalized meshes were removed and used for experiments.

Polypropylene meshes were cleaned similarly to steel and copper mesh. Cleaned meshes were then subjected to oxygen plasma treatment at 40% power for 2 min. They were then immersed in 200 mL of toluene containing 168 ppm water and 0.6 mL of trichlormethylsilane for 6 h. Following incubation, the meshes were rinsed thoroughly with n-hexane and dried in a vacuum oven at 49 °C for 1 h. The hydrophobic meshes were exposed to a second oxygen plasma treatment at 40% power for 2 min and immersed in 200 mL n-hexane containing 0.2 mL of perfluorodecyltrichlorosilane for 1 h. Finally, the meshes were rinsed thoroughly with fresh n-hexane and dried in a vacuum oven at 49 °C for 1 h. This sequential silanization process produced the fluorinated nanofilament coating on the mesh surface.

### b. Experiments

The experimental setup consisted of a drop dispensing needle, a superhydrophobic mesh, a Faraday cup, an electrometer and a high-speed camera (Fig. 1a). All experiments were conducted at 23 ± 2 °C temperature and 30-50% RH. First, drops of 25 µl deionized water (18.2 MΩ.cm) were generated using a peristaltic pump (Minipuls 3, Gilson). The drop, formed from a grounded syringe needle, was allowed to fall onto a grounded superhydrophobic mesh directly below it. Fig. 1(b) shows the SEM image of the superhydrophobic steel mesh used in present study with contact angle of 155° (Fig. 1c). Secondary droplets that passed through the mesh due to the impact were collected in a Faraday cup positioned directly beneath the mesh. An electrometer (6514 System Electrometer, Keithley) was connected to the Faraday cup to measure the charge. A high-speed camera (FASTCAM Mini AX, Photron, 10,000 fps, shutter speed 1/frame, resolution 768 × 528 pixels) was positioned directly beside the mesh to capture the dynamics of drop impact and atomization. The distance (i.e., height) between the mesh and the syringe needle was varied from 2 to 50 cm to control velocity. At each height three measurements were recorded. The total mass of all secondary droplets was determined by a microbalance (EX225D, OHAUS) for 5 consecutive drop impacts at a particular height. Drop dynamics was analyzed using ImageJ software. For very small secondary droplets, ImageJ was used to determine the size of the droplets and thus, mass.

## III. RESULTS AND DISCUSSION

The impact of a drop on a superhydrophobic surface proceeded through a series of distinct stages, each governed by the interplay of inertial, surface tension, and viscous effects.[16,19-23] Immediately upon contact with the mesh surface, the drop underwent rapid radial spreading driven predominantly by inertial forces (Fig. 1d, 0 to 4 ms). Due to the intrinsic porosity of the mesh, a considerable fraction of the drop penetrated through the open pores (Fig. 1d, 1 to 4 ms). After reaching its maximum spreading, the effect of inertia reduces, and surface tension becomes the dominant force. The surface tension force pulled the liquid back toward the center, causing the drops diameter to shrink. Since the mesh surface was superhydrophobic, drops retracted on the mesh without sticking. Finally, the stored surface energy is converted into upward motion, causing the droplet to bounce off and detach from the mesh surface (Fig. 1d, 24 ms).

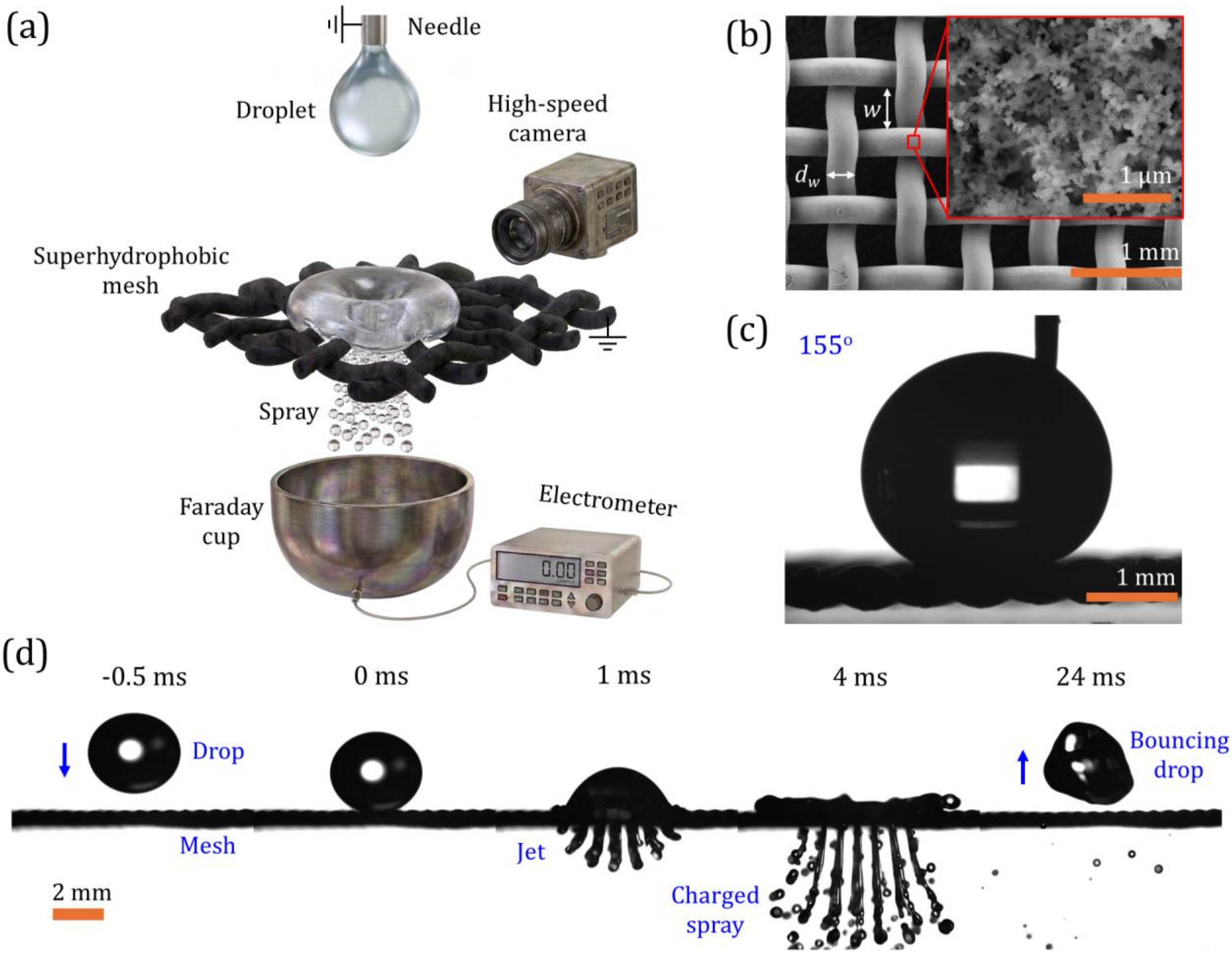

*Fig. 1: (a) Experimental setup for drop impact on superhydrophobic meshes and measurement of charge from spray. (b) SEM image of a steel mesh ($w$ = 400 µm, and $d_w$ = 330 µm) with superhydrophobic coating. Inset: structures on mesh at micro-nano scales. (c) Contact angle measurement for the superhydrophobic steel mesh. (d) High speed images of 25 µl drop impact on steel mesh at We ~ 100.*

We measured the charge of the ejected spray as a function of the Weber number. The Weber number quantifies the ratio between inertial and surface tension forces and is defined as $We = \rho U^2 D_0/\gamma$, where $\rho$ is the liquid density (997 kg/m$^3$), $U$ is the impact velocity, $D_0$ is the droplet diameter (~ 3.7 mm), and $\gamma$ is the surface tension of water (72 mN/m). The impact velocity was calculated with $U = \sqrt{2gh}$, where $h$ is the fall height before impact and $g$ is gravitational acceleration (9.81 m/s$^2$). As shown in Fig. 2(a), no measurable charge was detected at low $We$ (< 20). With increasing $We$, the charge carried by the spray increased, reached a maximum value at $We$ ~ 100. Beyond this $We$, the charge remained nearly constant and showed a saturation behavior, even when the $We$ was increased up to 500.

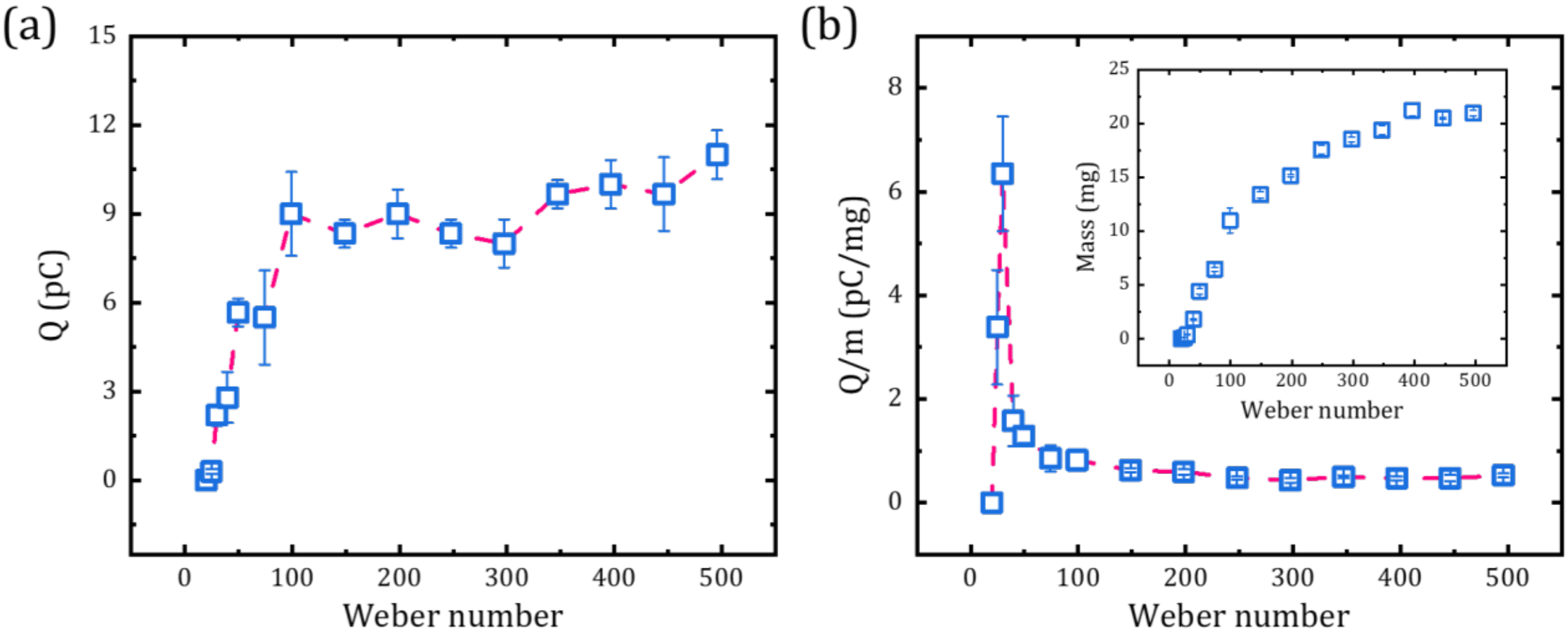

*Fig. 2: (a) Total charge (Q) of all the secondary droplets sprayed through a superhydrophobic steel mesh (w = 400 µm, and $d_w$ = 330 µm) versus Weber number. (b) Normalized charge (with respect to mass of the spray, Q/m) variation with Weber number. m is the total mass of all secondary droplets. Inset: spray mass with change in Weber number.*

The observed increase in spray charge with increasing $We$ may suggest that the enhanced charge ($Q$) originates simply from the increased amount of liquid ejected through the mesh. However, this interpretation is misleading. Although the mass of the generated spray increased with increasing $We$ and saturated above $We$~100 (Inset, Fig. 2b), the charge normalized by the spray mass ($Q/m$) exhibited a different trend. It increased to 6 pC/mg at $We$ = 30 and then decreased with increasing $We$. It eventually reached a nearly constant value of 0.5 pC/mg (Fig. 2b). This behavior indicates that the charge accumulation in spray is not governed solely by the amount of liquid participating in the spray formation process. The spray mass increased at a faster rate than the total charge carried by the spray. Therefore, the increase in charge with $We$ is not proportional to the amount of liquid ejected through the mesh. Instead, a progressively smaller fraction of the generated spray contributes to charge accumulation. Additionally, at lower $We$ (< 20), no charge was detected because of no spray formation.

To establish a correlation between the droplet dynamics and the measured spray charge, high-speed imaging was employed. As the charge generation is expected to originate from the interaction between the liquid and the mesh, particular attention was given to the evolution of the drop–mesh contact diameter ($D_c$). Fig. 3(a) shows the temporal evolution of the normalized contact diameter ($D_c/D_0$) for different $We$. At low $We$, the contact diameter increased during the inertial spreading stage and subsequently decreased gradually as the drop retracts under the action of surface tension. In contrast, at high $We$ (> 75), while the spreading phase remains relatively smooth, the retraction became markedly more abrupt. Following maximum spreading, the contact diameter decreased rapidly, indicating a faster withdrawal of the liquid from the mesh.

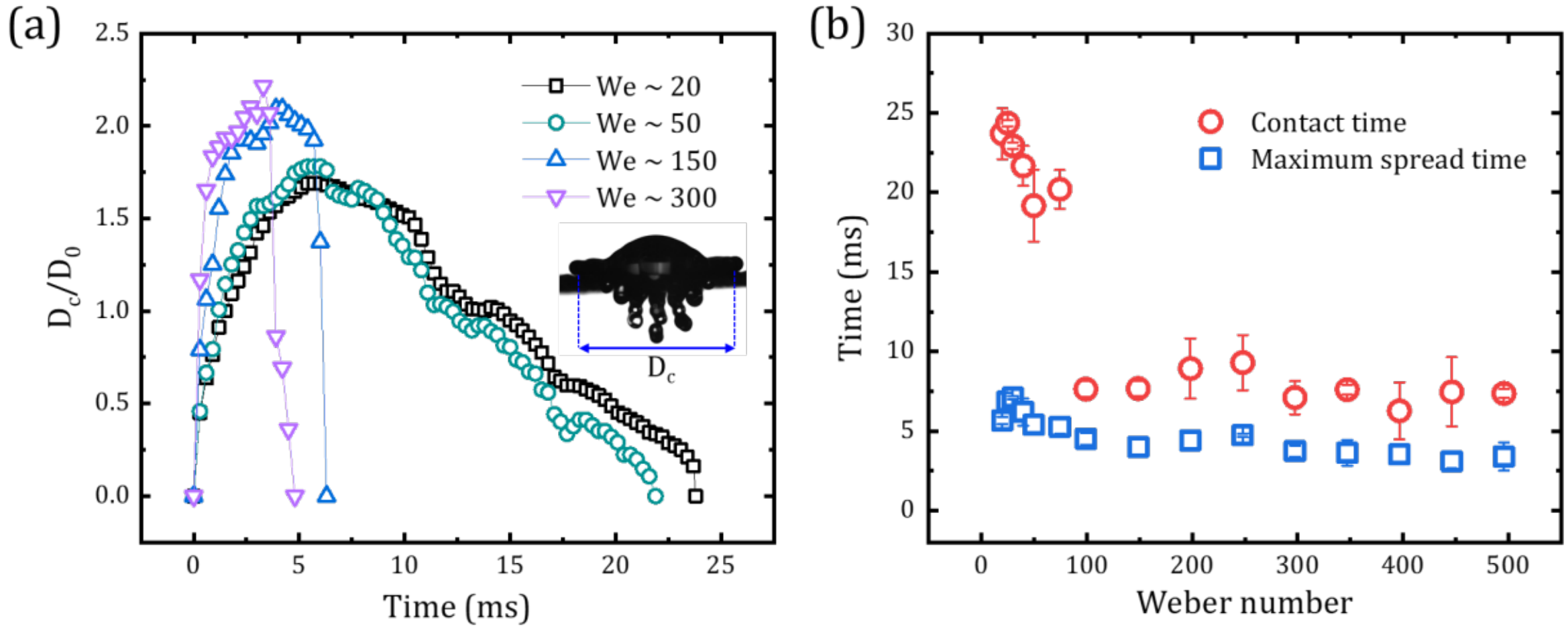

*Fig. 3: (a) Time evolution of the normalized contact diameter ($D_c/D_0$) for different Weber number impact on superhydrophobic steel meshes (w = 400 μm, and $d_w$ = 330 μm). Inset figure represents the measurement scenario of contact diameter. (b) Variation of contact time and maximum spread time of the drop on superhydrophobic steel mesh (w = 400 μm, and $d_w$ = 330 μm) with Weber number.*

To further quantify this effect, the maximum spreading and contact times were plotted (Fig. 3b). Contact time was defined as the time between initial contact of the drop to the time when the center of the rebounding drop loses contact with the mesh. At low $We$, the drop remains in contact with the mesh for a relatively long duration (~ 24 ms). Then, the contact time decreased rapidly with increasing $We$ and reached a nearly constant value of approximately 6 ms. In contrast, the time required to reach the maximum spreading diameter exhibited a weaker dependence on the $We$ (Fig. 3b). The reduction in contact time mirrors the trend observed for the spray charge (Fig. 2a). In this region, some of the penetrated liquid inside pores is rapidly pulled back due to jet pinch off, producing an upward momentum that lifts the droplet from the surface before complete lateral retraction. As a result, the droplet detaches in a flattened, pancake-like shape, leading to a substantially reduced contact time.[24-26]

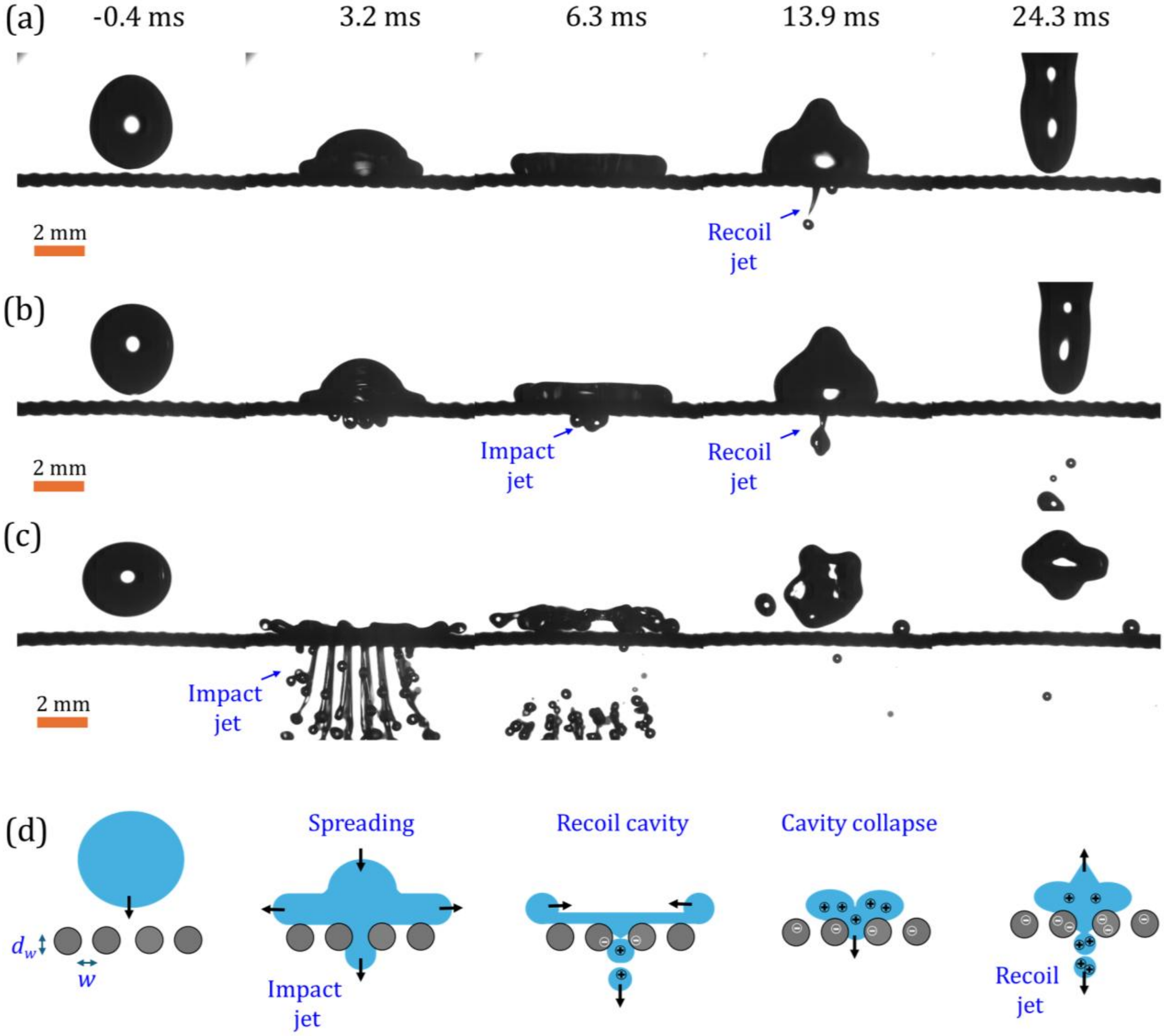

*Fig. 4: High speed images of 25 µl droplet impact on steel mesh (w = 400 µm, and $d_w$ = 330 µm) showing (a) recoil jet at We ~ 25, (b) impact and recoil jet combined at We ~ 30 and (c) only impact jet at We ~ 150. (d) Schematic representation of charging mechanism in the intermediate regime from impact and recoil jet.*

From the high-speed videos three distinct spray-generation regimes could be distinguished (Supplementary video S1, S2 and S3). At low $We$ (~ 25), spray was generated exclusively through recoil-jet formation (Fig. 4a, Supplementary video S1). At intermediate $We$ (30 < $We$

< 50), both recoil-jet and impact-jet mechanisms contributed to spray generation (Fig. 4b, Supplementary video S2). At high $We$ (> 50), spray was produced solely by impact-induced penetration (Fig. 4c, Supplementary video S3).

At low $We$ (= 25), the drop remained attached to the mesh after reaching its maximum spreading diameter and underwent a complete retraction cycle. Charge generation is expected to occur primarily during the separation of liquid and solid interfaces.[27,28] Thus, the primary drop acquires a positive charge while the contact line recedes. During retraction, cavity collapse drives liquid penetration through the mesh, producing a recoil jet.[16,29] Consequently, when the recoil jet penetrates through the mesh, it ejects liquid that is already charged (Fig. 4d). This provides a mechanism for the relatively high charge per unit spray mass observed at low $We$.

As the $We$ increases (~ 30), impact-induced penetration begins immediately upon drop impact. Liquid jets pass through the mesh pores during the impact phase (impact jet). Charge is generated as the liquid separates from the solid interface. Because the impacting drop is initially electrically neutral, the impact-jet contribution is limited to the charge generated locally within the pores (Fig. 4d). In addition, in this regime, recoil-driven penetration still occurs after maximum spreading. Therefore, both mechanisms operate simultaneously: the impact jet contributes locally generated charge, while the recoil jet additionally ejects liquid that has already acquired charge during droplet retraction. The coexistence of these two mechanisms explains the enhanced charge output in this regime.

At high $We$ (>75), the spray is produced before the contact line of the primary drop has receded. In this phase the primary drop is not charged. When pancake bouncing occurs, the rebound mode changes fundamentally. The droplet lifts off immediately after maximum spreading, before a recoil jet can develop (Fig. 4c showing $We$ ~ 150). Once contact with the mesh is lost, recoil-driven charging is minimized or suppressed. Consequently, spray generation is dominated solely by impact-induced penetration, and the additional contribution from the charged recoil jet disappears.

Spray generation by recoil-jet formation at low and impact-jetting at high $We$ determine the charging of the satellite droplets. This raises the question of which parameters describe the onset and transition between the two processes. Impact based liquid penetration occurs only when the impact-induced dynamic pressure overcomes the capillary pressure that resists liquid entry into the pores.[19,20] The impact pressure is governed by the dynamic pressure and scales as ~ $\rho U^2$.[19] The capillary pressure resisting penetration is determined by the surface tension force acting along the pore perimeter and scales as ~ $\gamma L / A$, where $A$ is the pore opening area ($w^2$), and $L$ is the perimeter ($4w$). Thus, for spray formation, $\rho U^2 \sim \gamma L/A$ should satisfy or $We_i \sim LD_0/A$. For 400 $\mu m$ pore size and 25 $\mu l$ ($D_0$ = 3.65 mm) droplet, $We_i \sim 37$ is required for impact spray formation, which matches very well with our observation ($We \sim 30$). The onset of recoil penetration occurs when recoil-induced pressure overcomes the resisting capillary pressure.[19,29] The critical recoil Weber number for penetration is given by $We_r \sim (LD_0/A)^{2/3}$.[19] For 400 $\mu m$ pore size and 25 $\mu l$ ($D_0$ = 3.7 mm) droplet, $We_r > 12$ is required for recoil jet formation. Which is very close to experimental value (> 20).

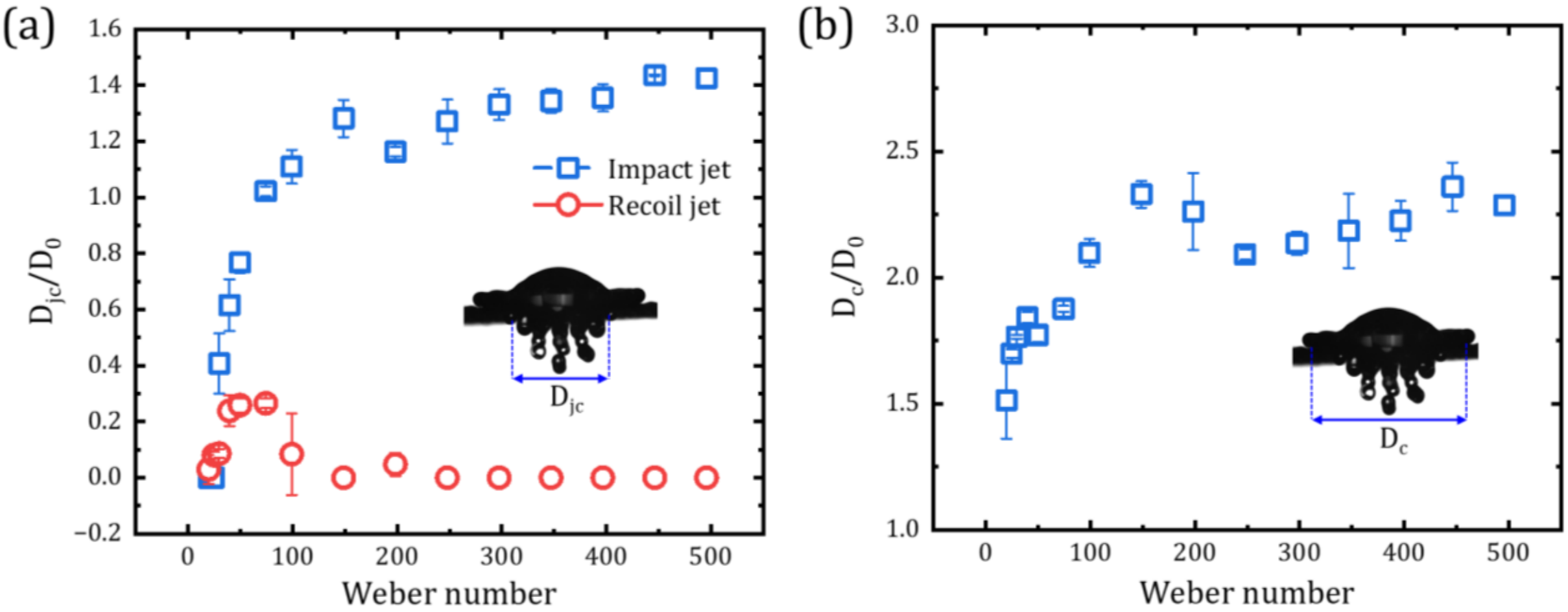

*Fig. 5: (a) Normalized jet contact diameter ($D_{jc}/D_0$) change with $We$. Inset represents the measurement snapshot. (b) Normalized droplet contact diameter ($D_c/D_0$) change with $We$. Inset represents the measurement snapshot. Both plots are for steel mesh ($w$ = 400 μm, and $d_w$ = 330 μm).*

The charge due to impact jet is primarily dominated by the separation at the pores. Thus, spray charge due to impact depends on the area from which jet is coming out ($D_{jc}$). The diameter of the jet contact initially increases with $We$ and then saturates (Fig. 5a). Thus, it is expected to see charge saturation at higher $We$ (Fig. 2a). The saturation in jet contact diameter is because of the droplet momentum redirection from perpendicular to mesh to the parallel to mesh.

The charge due to the recoil jet is dominated by the maximum spread contact diameter ($D_c$). Since the maximum spread contact diameter increases with $We$ (up to 100, Fig. 5b), charging due to retraction is also expected to increase (Fig. 2a). Additionally, higher $We$ impact may lead to more water penetration inside superhydrophobic nanostructures and further enhance charge separation.[30] However, a pure recoil mode is limited and combine recoil with impact mode is more prominent around 30 < $We$ < 75 (Fig. 5a).

For practical energy-harvesting of mesh-induced spray charging, the mesh geometry must be optimized to maximize charge separation per unit spray mass. The present results indicate that the recoil jet mode produces a substantially larger $Q/m$ ratio than the impact jet mode. Therefore, extending the $We$ range over which recoil-driven ejection occurs is desirable. Previous studies have shown that reducing the pore opening ($w$) promotes and prolongs recoil jet ejection.[29] In addition, a smaller pore opening increases the effective liquid–solid interaction area within the mesh, which can further enhance charge separation. Consistent with this expectation, decreasing the pore opening leads to an almost order-of-magnitude increase in the measured charge per unit spray mass (Fig. 6a). The enhancement is most pronounced for meshes that combine small pore openings with small wire diameters, since this geometry supports recoil-dominated spray formation while maintaining larger contact area.

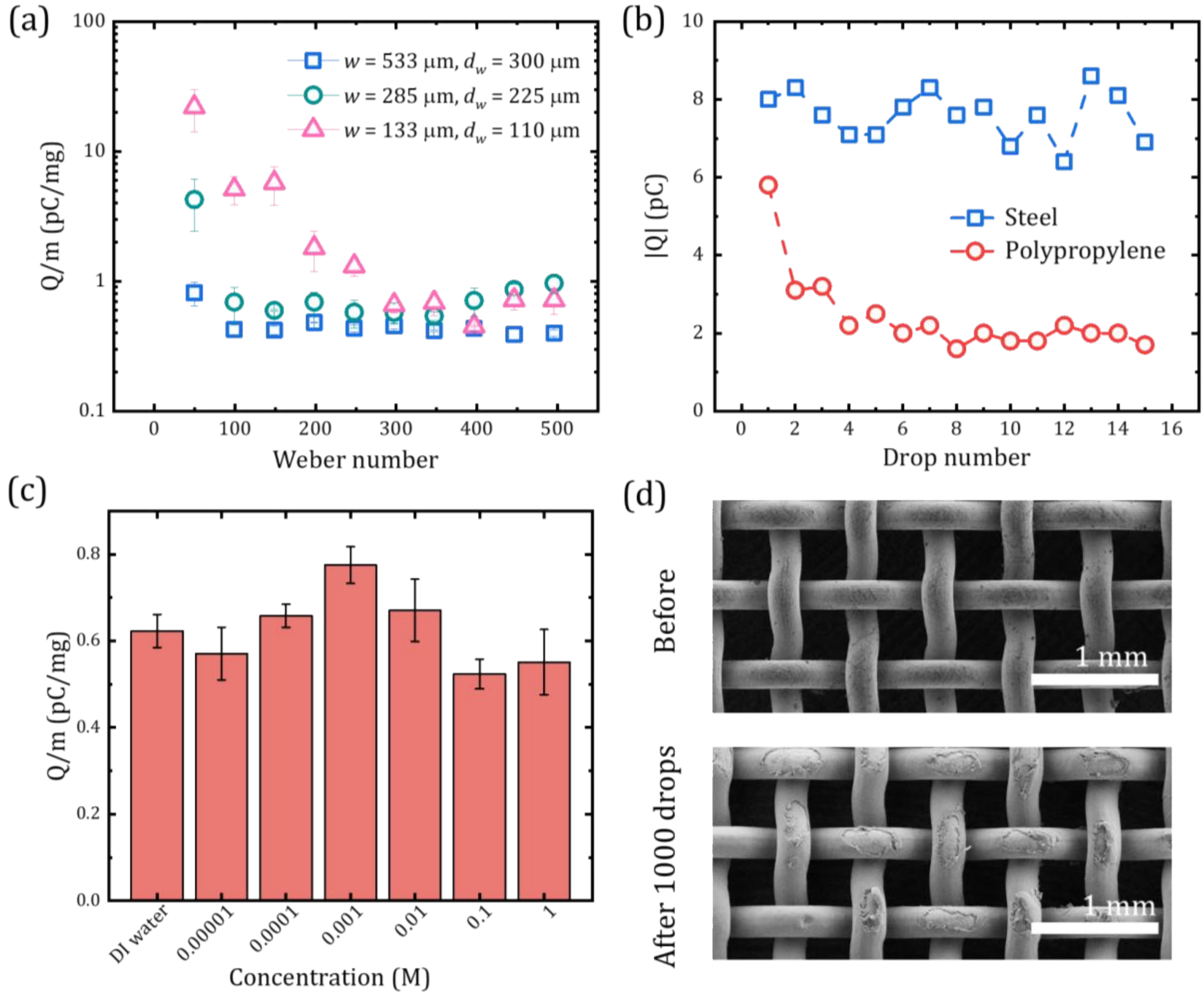

*Fig. 6: (a) Normalized charge (respect to mass of the spray, Q/m) variation with Weber number for different sizes of copper mesh* ($w$ = 533 μm, 285 μm and 133 μm pore sizes with $d_w$ = 300 μm, 225 μm and 110 μm wire diameter, respectively). *(b) Absolute charge for continuous droplet impact on Steel (w = 400 μm and $d_w$ = 330 μm) and polypropylene mesh (w = 500 μm and $d_w$ = 300 μm) at $We$ = 100. Time interval between two droplet was 2 s. (c) Charge dependence on various NaCl concentrations at $We$ ~ 150 for steel mesh (w = 400 μm and $d_w$ = 330 μm). (d) Steel soot superhydrophobic mesh (w = 400 μm and $d_w$ = 330 μm) before and after impact of 1000 drops.*

Because the charge generated by a single drop impact is small, practical energy harvesting requires operation at high drop-impact frequencies. Under such conditions, the charge dissipation of the mesh between successive impacts becomes important. If the surface retains residual charge for a long time, the next drop encounters a partially charged interface, which reduces the effective charge separation during liquid–solid interaction.[31] Therefore, meshes with slow charge-dissipation dynamics are not ideal for continuous operation. Using a conductive base mesh, such as metal, provides a rapid pathway for charge relaxation, allowing the surface potential to recover between impacts (Fig. 6b). As a result, each incoming drop interacts with an effectively refreshed surface, promoting repeatable and maximized charge separation. In contrast, insulating meshes such as polypropylene retain charge over longer timescales. This residual surface charge progressively screens further charge transfer, causing the charge generated per drop to decrease with repeated impacts and eventually saturate at a lower steady value after only a few droplets.

The effect of salt concentration (NaCl) on the charge separation on steel mesh at $We \sim 150$ is shown in Fig. 6(c). The charge output initially increased with increasing salt concentration, reaching a maximum value of 0.78 pC/mg at 1 mM, followed by a gradual decrease with further increase in concentration up to 1 M. Thus, the salt concentration can be modulated to enhance energy output. A similar enhancement in charge output was also observed for droplet sliding on hydrophobic surfaces.[32,33]

Although superhydrophobic surfaces are desirable because they enable rapid drop removal, they are often mechanically fragile. Repeated drop impacts can damage the surface microstructure (Fig. 6d), leading to deterioration of its superhydrophobic properties and changes in wettability, which may lead to change in spray charge.

# IV. CONCLUSION

This study shows that the impact of water drops on superhydrophobic meshes can produce charged sprays through liquid–solid contact separation inside and around the mesh pores. We identify four distinct regimes of spray charging, separated by three characteristic transition Weber numbers. In the first regime ($We < 20$), no liquid penetrates the mesh, and consequently no measurable charge is generated. In the second regime ($20 < We < 30$), recoil jets dominate spray formation, producing a relatively high charge per unit spray mass due to charge accumulation during droplet retraction. In the third regime ($30 < We < 75$), the combined action of recoil and impact jets enhances the total charge output. In the fourth regime (We > 75), pancake bouncing suppresses recoil-jet formation, and charging becomes dominated by impact-induced penetration. This leads to saturation of the total charge and a lower normalized charge.

Mesh geometry plays a key role: smaller pores increase the charge-to-mass ratio. Conductive meshes allow rapid charge dissipation and stable performance under repeated droplet impacts. These findings clarify the mechanisms of mesh-induced spray charging and provide guidelines for designing passive systems for droplet-impact-based charge separation and potential rain-energy harvesting.

## ACKNOWLEDGMENTS

The authors gratefully acknowledge Jürgen Thiel for help in making superhydrophobic mesh and Chitransh Upreti for assistance with the SEM imaging. The authors also thank Prof. Prosenjit Sen and Bheema Shankar Reddy for providing the mesh samples used in this study. Setup figure was partly generated using Google Gemini, followed by authors edit to ensure technical accuracy.

## SUPPLEMENTARY INFORMATION

Supplementary Video S1: High speed video of 25 $\mu$l droplet impacting on steel mesh at $We$ ~ 25. Video taken at 10000 fps and playing at 30 fps.

Supplementary Video S2: High speed video of 25 $\mu$l droplet impacting on steel mesh at $We$ ~ 30. Video taken at 10000 fps and playing at 30 fps.

Supplementary Video S3: High speed video of 25 $\mu$l droplet impacting on steel mesh at $We$ ~ 150. Video taken at 10000 fps and playing at 30 fps.

## AUTHOR DECLARATIONS

### Conflict of Interest

The authors have no conflicts to disclose.

### Author Contributions

H.J.B., M.T. and R.L. conceived the project. R.K. carried out experiments. R.L. designed experiments. R.K. and R.L. analysed the data. R.L. and H.J.B. supervised the project. R.L. wrote the manuscript. R.L., R.K., M.T and H.J.B. discussed results, reviewed and revised the manuscript.

## DATA AVAILABILITY

The data that support the findings of this study are available from the corresponding authors upon reasonable request.